\documentclass[10pt,conference]{IEEEtran}

\author{
	\IEEEauthorblockN{
		Ruchit Rawal\IEEEauthorrefmark{2}\IEEEauthorrefmark{4},
		Victor-Alexandru Pădurean\IEEEauthorrefmark{2},
		Sven Apel\IEEEauthorrefmark{5},
            Adish Singla\IEEEauthorrefmark{2}, 
		Mariya Toneva\IEEEauthorrefmark{2}
        }
	\IEEEauthorblockA{\IEEEauthorrefmark{2}Max Planck Institute for Software Systems, Saarbrücken, Germany}
 	\IEEEauthorblockA{\IEEEauthorrefmark{5}Saarland University, Saarbrücken, Germany} 
    \IEEEauthorblockA{\IEEEauthorrefmark{4}Corresponding Author} 
	    \IEEEauthorblockA{
	rawalruchit22@gmail.com, vpadurea@mpi-sws.org, apel@cs.uni-saarland.de, adishs@mpi-sws.org, mtoneva@mpi-sws.org}
 }

\usepackage{cite}
\usepackage{amsmath,amssymb,amsfonts}
\usepackage{algorithmic}
\usepackage{graphicx}
\usepackage{textcomp}
\usepackage{xcolor}

\usepackage{enumitem}
\usepackage{caption}
\usepackage{subcaption}
\usepackage{varwidth}
\usepackage{url}
\usepackage{placeins}
\usepackage{stfloats}
\usepackage{float}
\usepackage{tcolorbox}
\usepackage{hyperref}

\definecolor{customlightgray}{RGB}{237,237,237}

\newcommand{\change}[1]{#1}

\usepackage[normalem]{ulem}   
\usepackage{xcolor}           
\definecolor{reddish}{RGB}{200, 0, 0}

\usepackage{cleveref}
\def\BibTeX{{\rm B\kern-.05em{\sc i\kern-.025em b}\kern-.08em
    T\kern-.1667em\lower.7ex\hbox{E}\kern-.125emX}}

\begin{document}

\title{Hints Help Finding and Fixing Bugs Differently \\ in Python and Text-based Program Representations}

\maketitle

\begin{abstract}
With the recent advances in AI programming assistants such as GitHub Copilot, programming is not limited to classical programming languages anymore--programming tasks can also be expressed and solved by end-users in natural text. Despite the availability of this new programming modality, users still face difficulties with algorithmic understanding and program debugging. One promising approach to support end-users is to provide hints to help them find and fix bugs while forming and improving their programming capabilities. While it is plausible that hints can help, it is unclear which type of hint is helpful and how this depends on program representations (classic source code or a textual representation) and the user's capability of understanding the algorithmic task. To understand the role of hints in this space, we conduct a large-scale crowd-sourced study involving $753$ participants investigating the effect of three types of hints (test cases, conceptual, and detailed), across two program representations (Python and text-based), and two groups of users (with clear understanding or confusion about the algorithmic task). We find that the program representation (Python vs.\ text) has a significant influence on the users' \change{accuracy} at finding and fixing bugs. Surprisingly, users are more accurate at finding and fixing bugs when they see the program in natural text. Hints are generally helpful in improving \change{accuracy}, but different hints help differently depending on the program representation and the user's understanding of the algorithmic task. These findings have implications for designing next-generation programming tools that provide personalized support to users, for example, by adapting the programming modality and providing hints with respect to the user's skill level and understanding.
\end{abstract}

\begin{IEEEkeywords}
program comprehension, debugging, programming modalities, hints, crowd-sourced study
\end{IEEEkeywords}

\section{Introduction}
 Recent advances in generative AI, in particular, foundation models trained on text and source code, have the potential to make programming more accessible. Most notably, tools such as GitHub Copilot \cite{copilot} and ChatGPT \cite{ChatGPT} enable end-users to solve programming tasks through different modalities, including natural language text and pseudo-code specifications. Thus, programming is not limited to classical programming languages such as Python or C anymore, and programming problems can be expressed and solved by end-users in natural text. \change{This shift is particularly significant in the evolving landscape of software engineering, where new and diverse user groups are no longer relegated to peripheral roles but are increasingly taking a central role in developing applications -- many of which ($\sim\,60\%$) are now being created outside traditional IT departments by employees with limited or no technical development skills ($\sim\,30\%$) \cite{nocode_quixy}.}
 While these tools have enabled new forms of programming modalities, users still require algorithmic thinking and debugging skills to solve their programming problems~\cite{fitzgerald2008debugging,mccauley2008debugging,li2019towards,whalley2021novice}. Thus, there is a need to develop tools that can assist users with algorithmic understanding as well as finding and fixing bugs.

A series of recent works have explored how generative AI can be leveraged to support users with various forms of programming hints to help them find and fix bugs, while also forming and improving their programming capabilities~\cite{DBLP:conf/iticse/Prather00BACKKK23,DBLP:journals/corr/abs-2402-01580}. On the one hand, several works have proposed techniques that can provide tutor-style natural language hints~\cite{leinonen23sigcse,DBLP:conf/sigcse/WangMP24,DBLP:conf/icer/PhungPCGKMSS22,DBLP:conf/lak/PhungPS0CGSS24}; on the other hand, they also investigated how to design informative test cases for program comprehension and debugging~\cite{aaai24aied-kumar-testcases,edm24-heickal-ladders}. However, the role and merits of hints have been studied only for a classical programming setting, and it is unclear whether and how the helpfulness of hints depends on the program representation. Given the fact that AI programming assistants broaden the population of users, it is further open whether and how hints should be adapted to a user's skill level and degree of understanding of the algorithmic task.

\begin{figure*}[t!]
\centering

    \includegraphics[width=\linewidth]{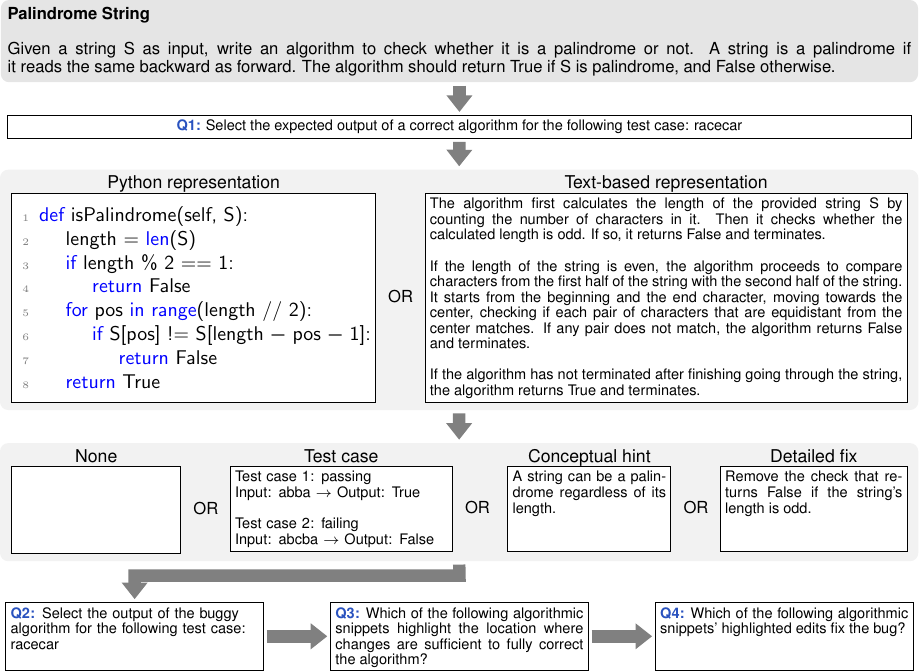}
     \caption{An illustrative example from the study showcasing an algorithmic task. After showing a task, the user is asked to answer a question related to understanding of the task (Q1). Afterward, the user is shown a buggy program (in Python or text-based representation), possibly along with a hint. Then, the user is asked to answer questions related to bug understanding (Q2), bug finding (Q3), and bug fixing (Q4). These questions are posed as multiple-choice questions--the answer options are not shown in the figure for brevity.
     }
    \label{fig.illustration}
\end{figure*}

\begin{figure*}[tp]
    \centering
    \includegraphics[width=0.9\textwidth]{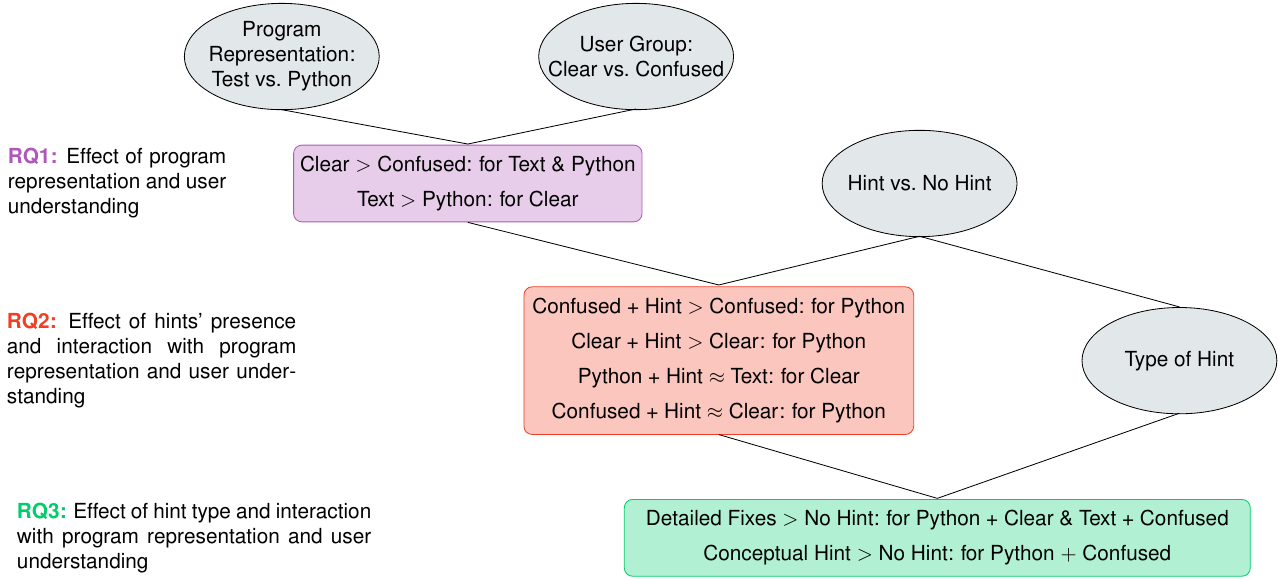}
    \caption{Visual summary of main findings to our Research Questions. Circular nodes represent main factors of variation (Program Representation, User Group, Hint Presence, and Hint Type). Rectangular blocks contain key takeaways, color-coded by research question (RQ1: purple, RQ2: red, RQ3: green). Connecting lines illustrate how factors combine to address different research questions.}
    \label{fig:rq_takeawyas}
\end{figure*}

In this paper, we study the interplay of hint types and program representations regarding their usefulness for end-users. More concretely, we investigate the effect of three types of hints (test cases, conceptual, and detailed), across two program representations (Python and text-based), and two groups of users (with clear understanding or confusion about the algorithmic task). We center our study around the following three research questions: 
\begin{itemize}[leftmargin=1cm,itemsep=1ex,topsep=1ex]
    \item[\textbf{RQ1}:] How does the program representation affect a user's ability to find and fix bugs, and how does this depend on the user's understanding of the task?
    \item[\textbf{RQ2}:] What is the utility of a hint on the user's ability to find and fix bugs across different program representations, and how does this depend on the user's understanding of the task?
    \item[\textbf{RQ3}:] What types of hint are more suitable for different program representations, and how does this depend on the user's understanding of the task?
\end{itemize}

To answer these research questions, we conducted a large-scale, crowd-sourced study involving $753$ participants. In this study, a participating user is presented with an algorithmic task along with a buggy program (in Python or text-based representation) and is asked to find and fix bug(s) in the provided program, possibly with the help of a hint. We illustrate the experiment flow in Figure~\ref{fig.illustration}. We measure the utility of a hint primarily in terms of a user's ability to successfully find/fix bugs through a set of multiple-choice questions; we also consider an increase in speed for accurate responses as a secondary indicator of helpfulness. We summarize the takeaways for each Research Question in Figure~\ref{fig:rq_takeawyas}. Our results for RQ1 show that the text-based program representation leads to better program debugging when considering users who demonstrate a clear understanding of the algorithmic task. Regarding RQ2, we found that hints for text representations 
\change{do not appear to be helpful in improving user accuracy}. On the other hand, hints for Python representations offer several advantages: (1) they improve user \change{accuracy} regardless of whether users have clear understanding or are confused; (2) they help reduce the \change{accuracy} gap between Python and text representations for users with a clear understanding; (3) they also reduce the \change{accuracy} gap between users who have clear understanding and who are confused about the algorithmic task.
Finally, when investigating the interplay of program representation and hint types for RQ3, we found that detailed fixes are generally the most helpful across all representation modalities and user types, and conceptual hints are particularly beneficial for users with a confusion about the algorithm task when working with Python representations.

Our results have implications for designing next-generation programming tools that can provide personalized support to users that is adapted to their experience and understanding. For instance, our results indicate that users with a clear understanding of the algorithmic task can benefit from text-based program representation. Moreover, our results showcase how user \change{accuracy} can be improved by providing different types of hints and adapting them according to the programming modality and the user's understanding of the task. 

Our main contributions are as follows: 
\begin{itemize}
    \item We analyze the helpfulness of various hint types for two different program representations.
    \item We conduct a large-scale study to investigate how different hint types and modalities affect a user's ability to find and fix bugs in a program.
    \item We examine how hint types and program modalities can be adapted to a user's understanding.
    \item We formulate a set of explicit hypotheses that are meant to inform further work in this field.
    \item We provide all raw data and analysis scripts: \url{https://github.com/bridge-ai-neuro/HintsCodeText}
\end{itemize}

\section{Related Work}

\subsection{Text vs.\ code for program comprehension}
\looseness-1Several studies have examined the cognitive processes that support a programmer's ability to understand programs written in code or natural language, finding shared \cite{liu2020computer,liu2024contribution} and distinct \cite{ivanova2020comprehension, floyd2017decoding} cognitive mechanisms, and differences in reading strategies \cite{busjahn2015eye}. Most previous studies have focused on passive comprehension paradigms where the participants were asked to simply read the program. A notable exception is Karas et al.~\cite{karas2021connecting}, who studied the functional connectivity in the brain during the writing of program code and prose. However, this study used different problems in the code and prose conditions, which makes it difficult to directly contrast the results. Our work complements these previous findings and focuses on the effect of program representation on finding and fixing bugs.

\subsection{Difference in program understanding for different users}
Most research on program comprehension does not consider inter-personal differences of programmers nor differences in their skills, levels of understanding the problem at hand, and backgrounds. In fact, most studies work with students as study subjects.
Notably, a line of research concentrates specifically on the effect of programming experience on program comprehension. For example, Uesbeck et al.\ \cite{USH+16} suggest that lambda expressions may hinder program comprehension, but only for novices, not experienced programmers. Stefik and Siebert\cite{SS13} found that syntactic constructs that are difficult for novices may guide teachers in choosing the appropriate language to start with. Burkhardt et al.\ \cite{BDW02} studied the effect of expertise on program comprehension. They found that the models experts build from a program and task differ from the models novices build, indicating different cognitive processes or mental modelling strategies involved. Vessey \cite{Ves85} studied the difference between experts and novices in debugging tasks. She found that experts adopt a holistic system view (and use breadth-first search), whereas novices do not (using depth-first search). In the same vein, D\'etienne \cite{Det02} notes that experts incorporate object-oriented and functional relationships in their reasoning, whereas novices focus on objects only. In a family of experiments, Dieste et al.\ \cite{DAU+17} found that years of experience are a suboptimal predictor of programmer performance, academic background and specialized knowledge of task-related aspects are better predictors. In contrast to these previous works, we define groups of users based on their understanding of the specific algorithmic task, which we measure empirically in the beginning of the same experiment.

\subsection{Supporting users with various forms of programming hints} A variety of assistive techniques have been considered in the literature that support users with programming hints with the goal of helping them find and fix bugs. Prior to recent developments in generative AI, automated techniques primarily focused on hints presented in the form of bug fixes because of challenges in automatically generating high-quality natural language hints~\cite{singh2013automated,DBLP:conf/pldi/GulwaniRZ18,yi2017feasibility,DBLP:conf/lats/HeadGSSFDH17}. Another line of research investigated crowd-sourcing approaches to obtain hints provided by other learners or tutors~\cite{DBLP:conf/lats/HeadGSSFDH17,hartmann2010would,al2018review}. Recent developments in generative AI have led to a surge in automated techniques that provide tutor-style natural language hints, for example, by providing conceptual hints without revealing details about fixes, thereby considering aspects of forming user's programming capabilities~\cite{leinonen23sigcse,DBLP:conf/lak/PhungPS0CGSS24}. Moreover, automated generative techniques have been proposed that provide effective error messages for syntactical errors~\cite{DBLP:conf/edm/PhungCGKMSS23,DBLP:conf/sigcse/WangMP24} or design informative test cases for a given buggy code~\cite{aaai24aied-kumar-testcases,edm24-heickal-ladders}. However, these works have considered hints only for classical programming settings, and it is unclear how the helpfulness of hints depends on programming modality, representations, and a user's skill level and capability to understand the problem at hand. Our work complements these works and focuses on understanding the interplay of hint types and program representations regarding their usefulness for end-users.

\section{Methodology}
  
\subsection{Experiment Design}
\label{subsec:exp_design}
To investigate how different program representations and hint conditions affect participants' ability to understand, identify, and fix bugs, we conducted a large-scale crowd-sourced study with a total of $753$ participants. We visualize the design flow of the study in Figure \ref{fig.illustration}. The study featured two primary types of program-representation conditions: text-based description and Python code. For each program representation type, there are four possible hint-conditions--no hint, test cases hint, conceptual hint, or detailed fix hint--resulting in a total of eight distinct conditions. Each participant was randomly assigned to one of these eight conditions. Within their assigned condition, participants were presented with two algorithmic tasks, each focusing on one of five different problems (see Section \ref{subsec:stimulus_design} for details about the problems).

Participants took an average of $15.76$ minutes to complete the survey. We recorded participants’ responses and response times for each question, as well as the time they spent on each step of the survey, including reading and understanding content on pages like the program page and the hint page. Before beginning the study, we asked participants to fill out a short demographics survey that included questions about their experience with programming, self-rated Python programming skills (on a scale of 1 to 10), and familiarity with English reading comprehension skills (on a scale of 1 to 10). Our participant pool predominantly consisted of individuals from English-speaking countries such as the USA, who rated themselves very highly on English reading comprehension. Regarding the programming questions, the participant pool was skewed towards individuals with less self-reported programming experience. We found the self-reported measures of programming experience and skills unreliable, as there was no strong relationship between these measures and the participants' understanding of the algorithmic task (i.e. accuracy on Q1); therefore, we decided not to use these self-reported measures for the main analyses, and instead group participants by the data-derived measure of Q1 accuracy, which indicates whether the participant correctly understood the problem description (see Sec. \ref{subsec:data_analysis} for more information about how this grouping was used in our analyses).

\subsection{Utility Metrics}
\label{subsec:utility_metrics}
We focus on two metrics to quantify the utility of program representation and the provided hints for program understanding and debugging:
\begin{enumerate}
    \item \textbf{Accuracy:} We assess participants' performance by computing the average accuracy over multiple-choice questions Q2, Q3, and Q4, as they cover different aspects of successfully debugging a program. 
    The range of average accuracy ranges from $0$ to $1$, with theoretical chance \change{accuracy} of $0.305$ (Q2: 4 options, Q3: 3 options, Q4: 3 options).
    \item \textbf{\change{Time Taken:}} We measure the average response time of participants when answering questions Q2, Q3, and Q4 correctly. \change{This includes the time taken to read the question, review the answer choices, and submit the response.} This metric allows us to determine whether accuracy gains come at the cost of efficiency (\change{inversely proportional to time taken}) or in addition to it.
    \end{enumerate}
By evaluating both accuracy and \change{time taken}, we gain a holistic understanding of how program representations and hints affect program debugging \cite{peitek2022correlates,middleton2024barriers}.

\subsection{Stimulus Design}
\label{subsec:stimulus_design}
Since we target end-users, we use $5$ basic computer science algorithmic tasks ranging in difficulty, commonly used in CS1 education, programming websites, and literature \cite{geeksforgeeks,DBLP:conf/icer/Fisler14,DBLP:conf/icer/PhungPCGKMSS22,DBLP:journals/corr/abs-2403-06050,bugspotter}. The problem titles with brief descriptions are as follows:
\begin{itemize}
    \item ``Sum Positive Values'' -- calculate the sum of the positive values in the input list A.
    \item ``Count NonNegatives and Negatives'' -- check whether there are more non-negative values than negative values in an input list A.
    \item ``Print Average Rainfall'' -- print the average of non-negative integers representing daily rainfall amounts in the input list A.
    \item ``Palindrome String'' -- check whether the input string S is a palindrome.
    \item ``Fibonacci to N'' -- print the list of numbers in the Fibonacci sequence till the input number N.
\end{itemize}

We have $5$ instances of buggy programs for each algorithmic task to ensure that our approach generalizes beyond specific implementations of the algorithmic task, leading to a total of $25$ program instances.
Figure~\ref{fig.illustration} shows an example of one instance of a buggy algorithm for the ``Palindrome String'' problem. We present both the Python and text-based representations for this instance. The bug is that the algorithm mistakenly treats all odd-length strings as non-palindromes. The figure also shows the three types of hints for this instance. \change{We provide examples of the other algorithmic tasks on our \href{https://github.com/bridge-ai-neuro/HintsCodeText}{GitHub repository}}. We provide details on how we constructed the Python and text representations of the programs, and the hints below:

\subsubsection{Python Representation-Condition}
\label{subsec:python_rep_con}
To obtain the buggy programs in Python representation for ``Palindrome String'' and ``Fibonacci to N'' problems, we took buggy Python attempts from recent literature used to benchmark generative AI models \cite{DBLP:conf/icer/PhungPCGKMSS22,DBLP:conf/lak/PhungPS0CGSS24} -- these are adapted from attempts publicly available on the platform \textsf{\small geeksforgeeks.com} \cite{geeksforgeeks}. Next, to obtain the buggy programs for ``Sum Positive Values'', ``Count NonNegatives and Negatives'', and ``Print Average Rainfall'', we manually plant bugs starting from the correct Python solution, based on the bugs we have encountered in our experience of working with students. As it is unclear how syntactic bugs can be reflected in text-based representations, we opt to include solely buggy codes with semantic bugs.
\change{Specifically, we first identified key sub-objectives necessary for correctly solving specific algorithmic tasks. For example, in the ``Print Average Rainfall" task, previous research \cite{DBLP:conf/icer/Fisler14} has highlighted several essential objectives: handling negative inputs (Negative), summing the inputs (Sum), determining the number of inputs (Count), addressing cases with zero inputs (DivZero), and calculating the average (Average). We then designed bugs that cause the program to fail in one or two of these sub-objectives. This approach ensures that the bugs target specific functional aspects of the task, leading to meaningful failures.}
Examples of \change{other} such bugs include wrongly initializing counter or accumulator variables, starting to iterate from index $1$ instead of index $0$, using wrong comparison logic in conditionals, and so on. The buggy codes we include can be fixed with a few localized changes. 

\subsubsection{Text Representation-Condition}
\label{subsec:text_rep_con}
For each Python program, we carefully craft a corresponding text-based representation that describes the Python program in natural language without using any programming concepts such as ``variables”, ``loops”, etc. These descriptions underwent several rounds of internal iterations to ensure accuracy, clarity, and faithfulness to the original Python code.

\subsubsection{Hint-Conditions}
\label{subsec:hint_cond}
As previously discussed, each participant is assigned one of four hint conditions: no hint, test cases hint, conceptual hint, and detailed fix hint. The hint conditions are carefully chosen to target different aspects of overall bug understanding, being grounded in providing support to students in programming education \cite{edm24-heickal-ladders,aaai24aied-kumar-testcases,DBLP:conf/icer/PhungPCGKMSS22}. The test case hint includes two input-output pairs, one representing a success case and the other a failure case, with minimal differences in their input. Conceptual hints highlight the underlying issue present in the program without suggesting specific changes, while detailed fixes focus solely on the necessary changes without explaining the underlying issue. Importantly, these hints are independent of the program representation and are applicable to both text-based and Python representations for a given problem type. The authors dedicated multiple iterations to handcrafting these hints, ensuring their clarity and applicability across both representations.

\subsection{Multiple-Choice Questions:}
\label{subsec:mcqs}
Our study consisted of four multiple-choice questions per problem type to assess participants’ understanding and evaluate the utility of hints. More specifically,

\begin{enumerate}[label=Q\arabic*.] 
\item \textbf{[Understanding of problem description]} \textit{Q. Select the expected output of a correct algorithm for the following test case: …?} This question had four options, with one correct answer and three distractors, resulting in chance \change{accuracy} of 0.25.

\item \textbf{[Understanding output of buggy program]} \textit{Q. Select the output of the buggy algorithm for the following test case: …?} This question also had four options, with one correct answer and three distractors, resulting in chance \change{accuracy} of 0.25.

\item \textbf{[Ability to localize bug]} \textit{Q. Which of the following algorithmic snippets highlight the location where changes are sufficient to fully correct the algorithm?} This question consisted of three options, each highlighting different portions of the program, with one correct answer and two distractors, resulting in chance \change{accuracy} of 0.33.

\item \textbf{[Ability to fix bug]} \textit{Q. Which of the following algorithmic snippets' highlighted edits fix the bug?} This question also had three options, each featuring highlighted modifications in the program, with one correct answer and two distractors, resulting in chance \change{accuracy} of 0.33.
\end{enumerate}

All questions remained consistent across different hint conditions. Q1 and Q2 were consistent across different program representation conditions, as they focused on the input/output behavior of an ideal and buggy program, respectively, which is consistent across both program representation conditions for a particular algorithmic task. Q3 and Q4 involved identifying and fixing bugs by selecting the option with the correct portion highlighted. These questions varied across different program representations; in the text-based condition, highlights were on the text stimuli, and in the Python condition, highlights were on Python stimuli. However, we ensured that the underlying content remained the same for both questions; the locations and fixes in the Python version corresponded to text descriptions in the text-based version. We provide the quantitative questions and answer choices for the algorithmic task example shown in Figure~\ref{fig.illustration} in our \href{https://github.com/bridge-ai-neuro/HintsCodeText}{GitHub repository}.

\subsection{Participant Recruitment:}
\label{subsec:part_rec}
Our experimental study was designed and implemented using Qualtrics. We recruited participants via Cloud Research, a crowdsourcing platform that builds on top of Amazon Mechanical Turk, but applies a series of stringent filters and quality control measures to significantly enhance the reliability of the participant pool. Our study received formal approval from the Ethics Review Board of our institute, ensuring that all ethical considerations were met. Before beginning the tasks, each participant provided informed consent. The study was structured so that each participant was expected to complete it within $30$ minutes. Participants were compensated at an hourly rate of $12$ USD.

\subsection{\change{Participant Demographics}}
\change{Our study was conducted over two weeks and  included a total of 753 valid respondents. 417 participants were male, 322 were female, 10 identified as non-binary, and the remainder preferred not to disclose their gender. The median age of participants was 39, with ages ranging from 20 to 78. The average programming experience in the past five years was 1.13 years, with a minimum of 0 years and a maximum of 5 years. The average self-rated Python programming skill level was 2.36, on a scale from 1 to 10.}

\subsection{Data Analysis:}
\label{subsec:data_analysis}
As described in Section~\ref{subsec:exp_design}, we asked each participant to complete two independent algorithmic tasks, each involving a different problem but the same type of program representation and hint.
Additionally, we applied a 2-second threshold per question to flag and exclude responses indicative of guessing, \change{based on a conservative estimate from an internal pilot survey. Of the 759 responses initially collected, 6 participants were excluded. The remaining responses had an average response time of 124 seconds and a median of 89.6 seconds.}

For our data analysis, participants were first grouped according to their understanding of the problem description, as evidenced by their accuracy on Q1: participants who answered Q1 correctly were labeled as the “clear” understanding group, while those who answered incorrectly were labeled as the “confused” understanding group. 
\change{To ensure our dataset consisted of independent observations and minimized within-person variance, we averaged the scores of the quantitative questions (Q2, Q3, Q4) across both task responses for participants consistently classified as either ``clear'' or ``confused'' in both tasks. If a participant fell into different groups across tasks, we randomly selected one of the two responses and discard the other response from all the analyses.
Similarly, for average time results, as discussed in Section III.B, we only focus on participants from the ``clear” group and only include responses where Q2, Q3, and Q4 were all correctly answered. For each participant, if Q2, Q3, and Q4 were correctly answered in both tasks, we averaged the response times across the two tasks. If these questions were correctly answered in only one task, we used the time from that task.
Importantly, all other conditions -- such as hint type and program representation -- were consistent across the two tasks based on our study design. This process ensured that our dataset includes only one data point per participant, aligning with the assumptions required for the statistical tests we will discuss shortly.}
Participants were also grouped based on the program representations they were assigned (Python vs.\ text). When analyzing the effect of hints, we aggregated over the three different hint types to compare the presence of hints to the absence of hints in RQ2. In RQ3, we analyzed the different hints separately in comparison to the no-hint condition. \change{For each RQ, we present bar plots of the mean utility metrics, with error bars representing the standard error of the mean.}

\subsection{\change{Significance Testing}}
We use the Wilcoxon rank-sum test \cite{wilcoxon1992individual} to determine significant differences between different conditions and indicate significant differences by asterisks in all result figures ($*$ for p-value $<0.05$, $**$ for p-value $<0.01$).
\change{The Wilcoxon rank-sum test relies on two key assumptions: the samples must come from populations with the same shape, and the observations must be independent. To ensure these assumptions are met, we test whether the samples share the same shape using the Kolmogorov-Smirnov test \cite{massey1951kolmogorov} before drawing any conclusions about statistical significance. Additionally, as we discussed earlier, we use only one data point per participant, which maintains the independence of observations.}
We use the Wilcoxon signed-rank test \cite{wilcoxon1992individual} to determine whether a particular condition is significantly different from the corresponding chance accuracy. We chose these statistical tests because they do not make assumptions about the underlying data distribution (e.g. that data is distributed according to a Normal distribution).
\change{Additionally, we use Cohen's d \cite{cohen2013statistical} to measure effect sizes for significant differences between conditions. We report Cohen's d and the difference of means in Section~\ref{sec:results}, along with any mention of significant differences between the two conditions.}

\section{Results}
\label{sec:results}

\begin{figure}[tp]
    \centering
    \includegraphics[width=0.45\textwidth]{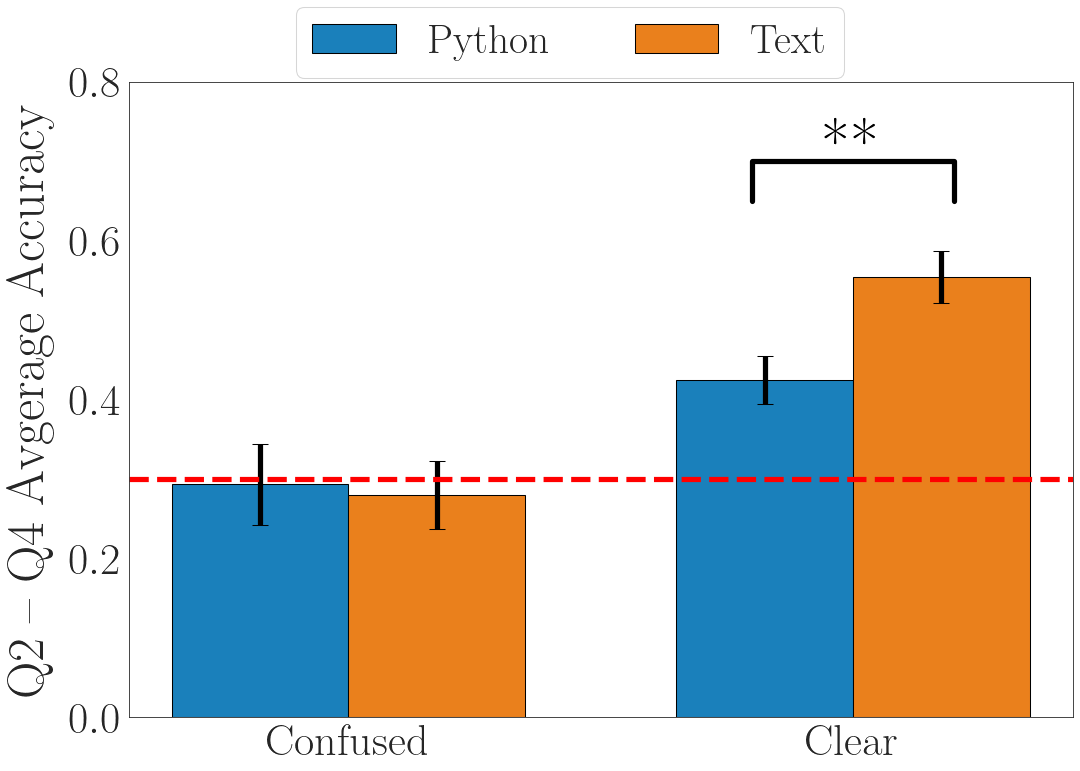}
    \caption{Accuracy of participants when presented with text-based vs. Python-based program representations and no hints. \change{The bar plot represents the mean Q2--Q4 average accuracy, with the vertical lines indicating the standard error of the mean.}
    The red dotted line represents chance \change{accuracy}, and significant differences between program representations are indicated with an asterisk ($*$).
    Surprisingly, clear participants perform significantly better when presented with text-based representations than Python-based representations.}
    \label{fig:eff_pres_mod}
\end{figure}

\textbf{RQ1.} \textit{How does program representation affect a user's ability to find and fix bugs, and how does this depend on the user’s understanding of the task?}

To investigate this question, we first compare the accuracy of participants who view the program in natural text with those who view it in Python. For this analysis, we only consider responses from participants in the ``No Hint” condition to avoid any potential interaction effects between the hint type and program representation. We further divide the analysis according to two groups of participants: those with a confused understanding and those with a clear understanding. The results are reported in Figure~\ref{fig:eff_pres_mod}.

Unsurprisingly, participants with a confused understanding of the problem description performed at chance levels (i.e., $0.305$) for both text-based and Python representation conditions. In contrast, participants with a clear understanding of the problem description performed significantly above chance in both the text-based \change{(p-value: $3.01 \times 10^{-9}$, effect-size: $0.959$, difference of means: $0.248$)} and Python representation conditions \change{(p-value: $3.75 \times 10^{-5}$, effect-size: $0.505$, difference of means: $0.119$)}. Additionally, these participants performed significantly better when viewing the program in text-based representations compared to Python-based representations (p-value: \change{${0.004}$, effect-size: $0.520$, difference of means: $0.129$)}.

\begin{figure*}[htp]
   \centering
   \includegraphics[width=0.9\linewidth]{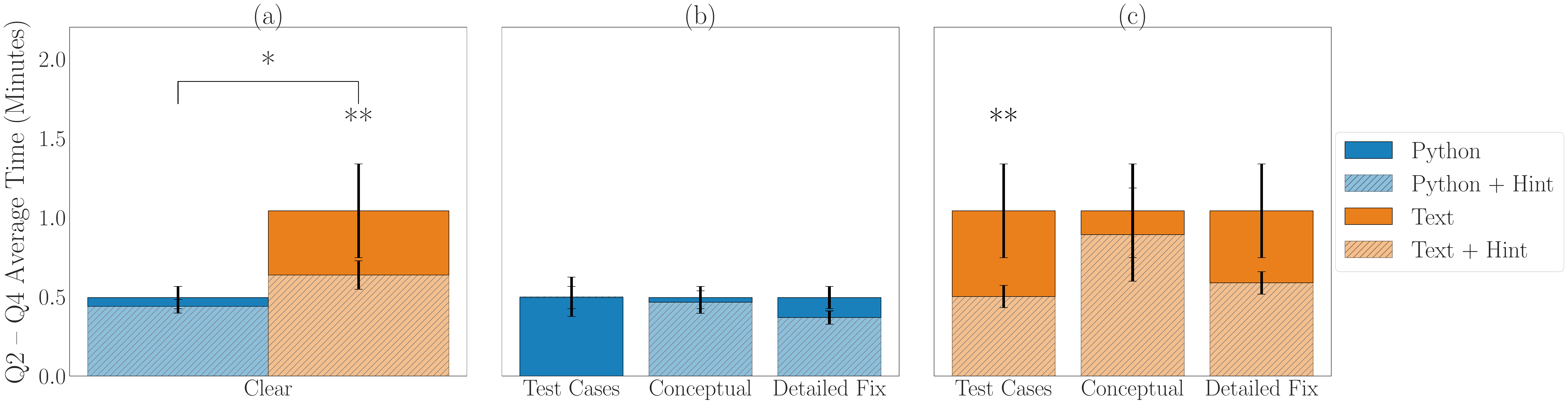}
   \caption{Time taken by participants to correctly answer Q2, Q3, and Q4 when presented with different hint types or no hint, across different program representations for clear understanding group participants. 
   \change{The bar plot represents the mean of the average time required to answer Q2--Q4 correctly, with the vertical lines indicating the standard error of the mean.}
   In (a), ``+ Hint" represents the average of the three different hint conditions, while in (b) and (c), we plot the different hint types separately. Significant differences between the no hint and hint conditions are represented by ($*$). Even though the hints may not be helpful in improving accuracy for the Text + Clear condition, they \change{help reduce} participants' \change{response time}. Similarly, improvement in accuracy doesn’t necessarily imply \change{reduction in response time} (Python + Clear condition). For the Text + Clear condition, \change{test cases significantly reduce response time}.}
   \label{fig:1x3_time_plot}
\end{figure*}

We further examine how program representation affects the participants' \change{response time}. To investigate this, we analyze the time taken by participants to correctly answer Q2 through Q4 in Figure~\ref{fig:1x3_time_plot}-(a). For this analysis, we only considered the time taken by participants who answered all three questions (Q2, Q3, and Q4) correctly. Moreover, we focused on the clear group participants, as only a few participants with a confused understanding remained after filtering out the incorrect responses. 
In Figure~\ref{fig:1x3_time_plot}-(a), we compared the average time taken by participants using text-based representations with Python representations and \change{find a text-based representations to take a significantly longer time (p-value: $0.047$, effect-size: $0.439$, difference-of-means: $0.547$).}
However, it is important to note that the average response time for questions is influenced by the time required to read program representations, as Q3 and Q4 multiple-choice options contain the original or modified program representations. To isolate this factor, we compared the 
\change{average time} for \change{only correctly answering} Q2, which has identical questions and multiple-choice options for both Python and text-based representations. We found no significant differences in response times for Q2 (\change{p-value \textgreater~$0.05$}).

\begin{tcolorbox}[
    colback=customlightgray,
    colframe=customlightgray,
    boxrule=0pt,
    arc=0pt,
    left=5pt,
    right=5pt,
    top=5pt,
    bottom=5pt,
    boxsep=0pt,
    width=\columnwidth
]
\textbf{RQ1 Takeaways:} For participants with clear understanding, text-based representations led to significantly better \change{accuracy} than Python representations. Confused participants performed at chance levels for both representations. 
\change{While text-based representations required more time overall, this was likely due to longer reading times rather than reduced efficiency in problem-solving.}
\end{tcolorbox}

\begin{figure}[htp]
    \centering
    \includegraphics[width=0.45\textwidth]{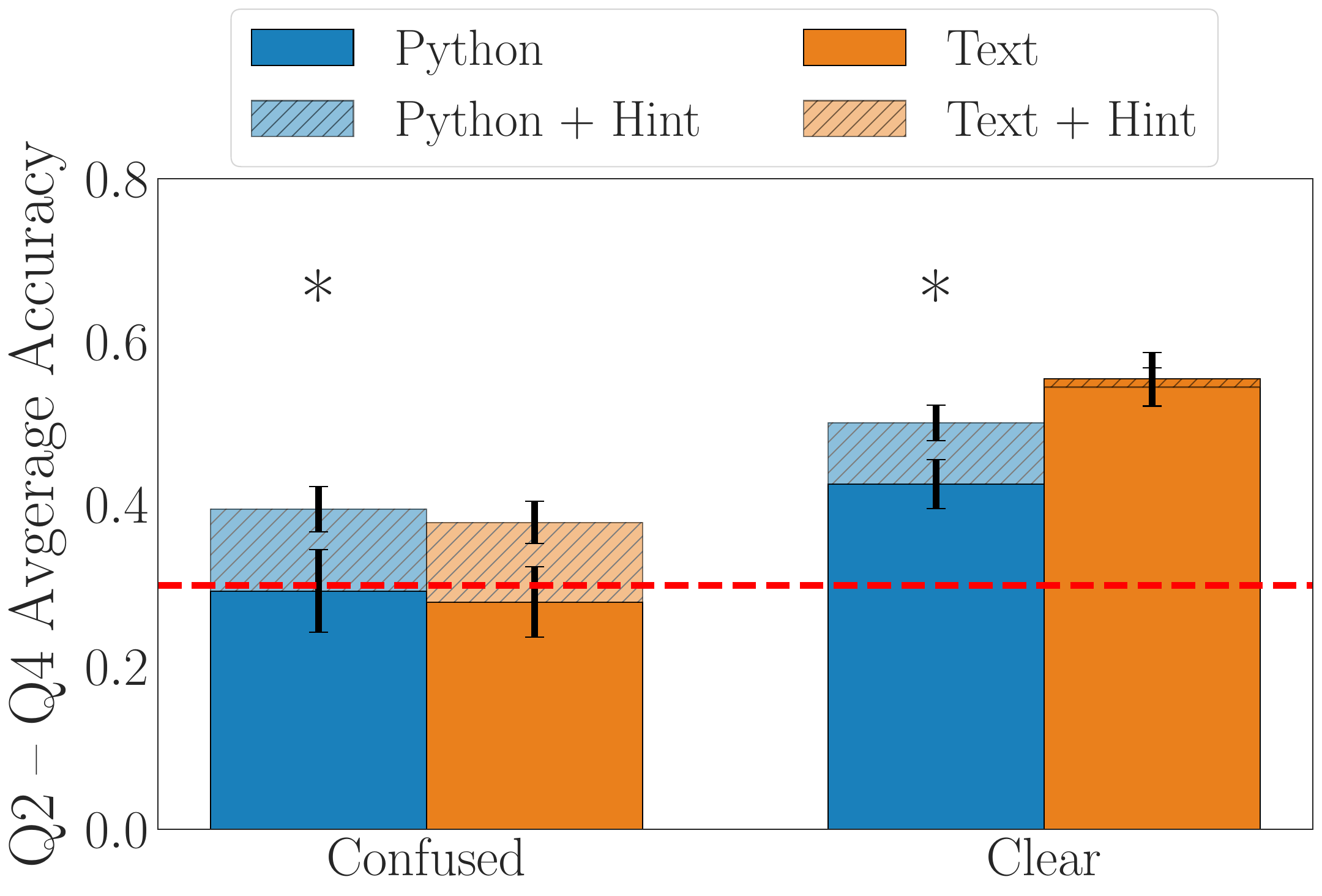}
    \caption{Accuracy of participants when presented with hints and no hints, across different program representations. \change{The bar plot represents the mean Q2--Q4 average accuracy, with the vertical lines indicating the standard error of the mean.}
    The red dotted line represents chance \change{accuracy}, and significant differences between the no hint and with hint conditions are indicated with an asterisk ($*$). Hints significantly improved \change{accuracy} for confused \change{and clear participants for Python program representations.} Hints also bridged \change{accuracy} gaps between representations (for clear participants) and understanding levels (for Python representation).}
    \label{fig:part_lbl_stim_und_acc_no_hints_vs_hints}
\end{figure}

\textbf{RQ2}: \textit{What is the utility of a hint on the user's ability to find and fix bugs across different program representations, and how does this depend on the user’s understanding?}

We investigate this question by comparing the \change{accuracy} of participants in the ``No Hint" condition with the average \change{accuracy} across the other three hint conditions, denoted via ``+ Hint" in Figure~\ref{fig:part_lbl_stim_und_acc_no_hints_vs_hints}. Given the significant effects of representation condition and participant understanding group observed in Figure~\ref{fig:eff_pres_mod}, we analyzed the utility of hints separately for text-based and Python representation conditions and for both clear and confused understanding groups of participants.

In Figure~\ref{fig:part_lbl_stim_und_acc_no_hints_vs_hints}, we present the accuracies when presented with any hint compared to the no hint condition, for each participant group and representation-condition pair. Significantly better \change{accuracy} with hints over no hint is indicated with an asterisk ($*$). For the Python representation condition, hints significantly helped both confused \change{(p-values 0.043 effect-size: 0.36, difference of means: 0.10)}
and 
clear participants \change{(p-value: 0.049, effect-size: 0.27, difference of means: 0.07}). In contrast, for text-based representation,
\change{while both clear and confused participants don't seem to be significantly helped by the provided hints, the confused participant groups show a small to medium effect size of 0.37 (p-value \textgreater~$0.05$, difference of means: 0.09).}

Confused participants, who initially performed at chance, improved significantly with hints in the Python condition. Additionally, the \change{accuracy} difference between the Python and text-based conditions for clear participants without hints disappeared when hints were provided for the Python condition \change{(p-value \textgreater~$0.05$, effect-size: 0.19, difference of means: 0.05)}. This suggests that hints help close the \change{accuracy} gap between different programming representations. Furthermore, for the Python condition, hints boosted the \change{accuracy} of confused participants to match that of clear participants without hints \change{(p-value \textgreater~$0.05$, effect-size: 0.116, difference of means: 0.031)}, showing that hints can reduce the gap between participants with varying levels of understanding.

\looseness-1In Figure~\ref{fig:1x3_time_plot}-(a), we analyze how hints affect the participants' \change{response time}. We note that hints improve efficiency (reduce \change{response time}) for text-based program representations \change{(p-value: 0.004, effect-size: 0.401, difference of means: 0.405)}, while for Python program representations, there is no significant difference between using hints and not using them. This result is complementary to the findings in Figure~\ref{fig:part_lbl_stim_und_acc_no_hints_vs_hints}, where hints improved the accuracy of clear group participants only for Python representations, with no significant effect on text-based representations.

\begin{figure*}[htp]
    \centering
    \includegraphics[width=0.725\linewidth]{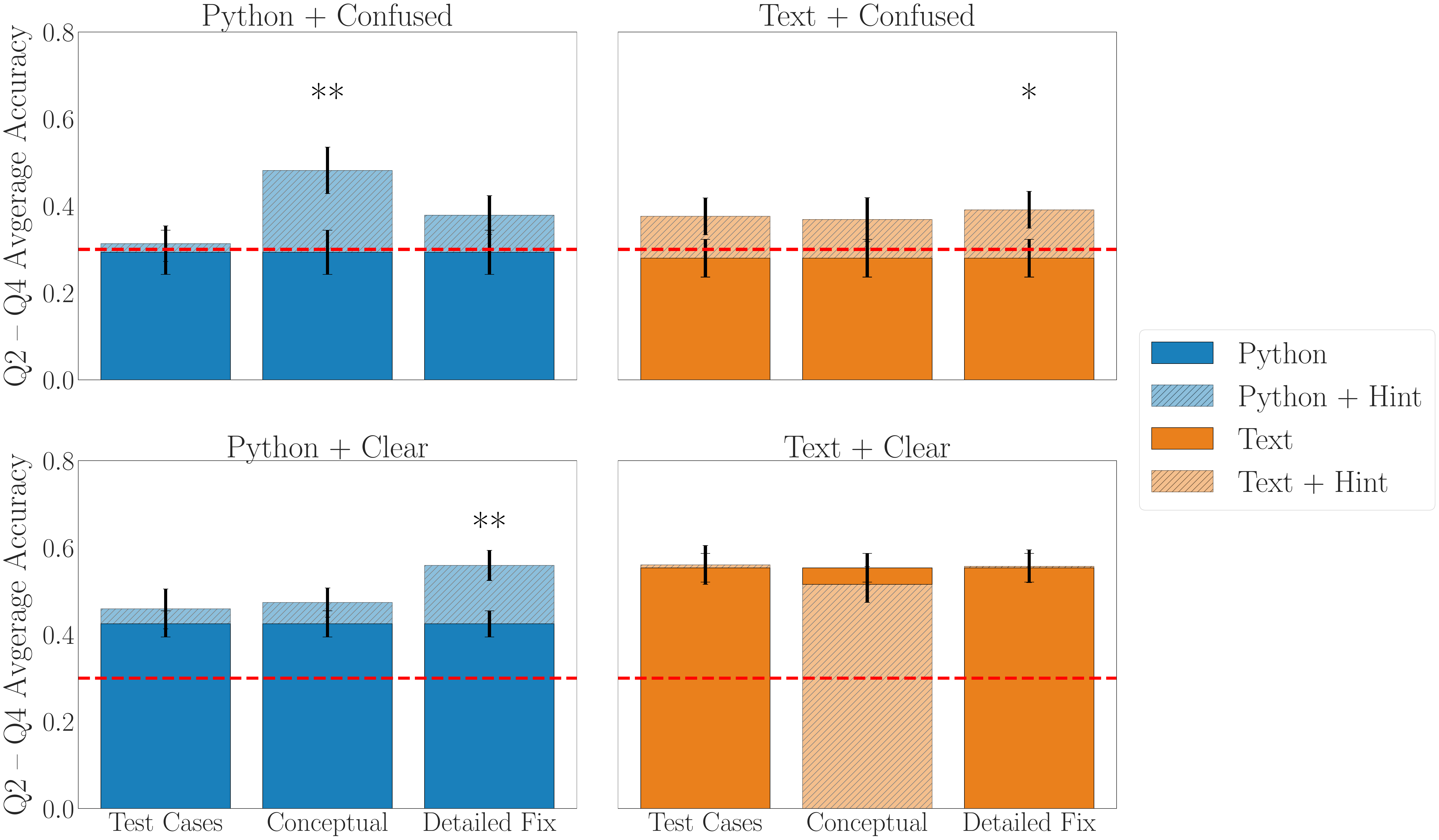}
    \caption{Accuracy of participants when presented with different hints or no hint, across different program representations and participants' level of understanding separately. \change{The bar plot represents the mean Q2--Q4 average accuracy, with the vertical lines indicating the standard error of the mean.}
    The red dotted line represents chance \change{accuracy}, and significant differences between the no hint and different hint type conditions are indicated with an asterisk ($*$). Detailed fixes are generally most helpful, while conceptual hints are particularly useful for participants with confused understanding in the Python representation condition.}
    \label{fig:interaction_eff}
\end{figure*}

\begin{tcolorbox}[
    colback=customlightgray,
    colframe=customlightgray,
    boxrule=0pt,
    arc=0pt,
    left=5pt,
    right=5pt,
    top=5pt,
    bottom=5pt,
    boxsep=0pt,
    width=\columnwidth
]
\textbf{RQ2 Takeaways:} Hints significantly improved \change{accuracy} for Python representations across both clear and confused understanding groups. For text-based representations, hints reduced \change{response time} for those with clear understanding. Additionally, hints bridged \change{accuracy} gaps between different representations and understanding levels.
\end{tcolorbox}

\textbf{RQ3:} \textit{What types of hint are more suitable for different program representations, and how does this depend on the user's understanding?}

As previously mentioned, we aggregated results over three different hint types for investigating \textbf{RQ2}. In this section, we explore whether different hint types are more beneficial for different program representations and whether this varies based on the participants' understanding. To investigate this, we extend our previous findings by analyzing the \change{accuracy} for each hint type separately and comparing whether they help improve \change{accuracy} compared to the no hint condition.

In Figure~\ref{fig:interaction_eff}, we present results for each hint type. 
Among the three types of hints, detailed fixes are generally the most helpful. They significantly improve \change{accuracy} for \change{clear participants with Python representations (p-value: 0.006, effect-size: 0.527, difference of means: 0.134), and confused participants with text-based representations (p-value: 0.047, effect-size: 0.484, difference of means: 0.111).}
Additionally, conceptual hints are especially beneficial for confused participants working with Python representations \change{(p-value: 0.009, effect-size: 0.626, difference of means: 0.188)}. We did not observe any significant effect of test case-based hints on \change{accuracy} compared to no hints. These analyses show that the effectiveness of hints varies depending on the program representation and the user's understanding level.

In Figure~\ref{fig:1x3_time_plot}-(b) \& -(c), we visualize the time taken by clear group participants to correctly answer Q2, Q3, and Q4. Consistent with our results while investigating RQ2, we note that \change{response time reductions} are only observed for the text-based program representations, where \change{test cases reduce the response time significantly (p-value: 0.002, effect-size: 0.512, difference of means: 0.541). We do not see significant effects for other hint types, and observe small effect-size for conceptual hint and (effect-size: 0.104), and small-to-medium effect size for detailed fix (effect-size: 0.408).}
\change{Notably, test cases did not improve the accuracy for any of the program-representation and participant-understanding conditions (refer to Figure~\ref{fig:interaction_eff})}.

\begin{tcolorbox}[
    colback=customlightgray,
    colframe=customlightgray,
    boxrule=0pt,
    arc=0pt,
    left=5pt,
    right=5pt,
    top=5pt,
    bottom=5pt,
    boxsep=0pt,
    width=\columnwidth
]
\textbf{RQ3 Takeaways:} Different hint types vary in utility depending on program representation and participant understanding. Detailed fixes generally improved \change{accuracy} most, while \change{test cases reduced response time} for participants with clear understanding viewing text-based representations.
\end{tcolorbox}

\section{Threats to Validity}
\label{sec:threats_to_val}
\subsection{Internal Validity} 

Internal validity refers to the ability to accurately establish cause-and-effect relationships between independent and dependent variables. In our study, internal validity is threatened by the use of crowdsourcing for data collection, which can introduce participant-dependent confounders. To mitigate this threat, we implemented a comprehensive randomization procedure that ensures that participants are randomly assigned to one of our eight distinct conditions. With more than 750 participants, we expect the influence of individual differences to be evenly spread across all conditions, effectively balancing out. Furthermore, we carefully considered the various independent variables during data analysis. For instance, when determining the effect of program representation (refer to \textbf{RQ1} in Section~\ref{subsec:data_analysis}), we analyzed only the ``No Hint"  condition to avoid confounding effects from the presence of hints. Likewise, we analyzed different hint conditions separately for each program representation. These methodological considerations strengthen the internal validity of our findings.

\subsection{Construct validity}
A key consideration in our study was ensuring construct validity, that is, reliably measuring the utility of different program representations and hints in bug understanding. As common in the literature, we measure utility through \change{accuracy} and \change{response time} metrics. We expect that a deeper understanding of a program and bug will lead to improved \change{accuracy} and reduced response time. To assess \change{accuracy}, we crafted three questions for each algorithmic task, probing participants' comprehension of the bug's impact (Q2), location (Q3), and potential fix (Q4). Please refer to Section~\ref{subsec:mcqs} for a detailed discussion regarding the different questions. The mean accuracy across these questions served as our \change{accuracy} metric. For \change{response time}, we measured the average time taken to answer these questions by participants who answered accurately. 
The construct validity for average time can be threatened by program representation. Specifically, text representations generally require longer reading times compared to Python code, and two of our questions (Q3 and Q4) included program representations in their answer choices. This could lead to artificially inflated response times for participants in the text condition, potentially masking true differences in \change{response time}. Therefore, while reporting the \change{response time} comparisons between Python and text representation conditions, we also report the \change{response time} based solely on Q2, which is exactly the same across both representations.

\subsection{Statistical conclusion validity}
To ensure statistical conclusion validity -- the degree to which conclusions we reach from analyses are accurate and appropriate -- we implemented several key measures in our study design and analysis. First, we secured a large sample size of $753$ participants, significantly enhancing the power of our statistical tests and allowing us to detect even subtle effects with sufficient confidence. In our analysis, we employed appropriate statistical tests that do not make assumptions about the underlying data distribution. Specifically, we examined the statistical difference between conditions using the Wilcoxon rank-sum test and the statistical difference between a condition and the theoretical chance \change{accuracy} using the Wilcoxon signed-rank test. 
\change{The Wilcoxon rank-sum test relies on two key assumptions: the samples must come from populations with the same shape, and the observations must be independent. To ensure these assumptions are met, we verify that the samples share the same shape using the Kolmogorov-Smirnov test \cite{massey1951kolmogorov} before drawing any conclusions about statistical significance. Additionally, since we use only one data point per participant, the independence of observations is maintained.}
We indicate significant differences in all result figures using asterisks ($*$ for p-value $<0.05$, $**$ for p-value $<0.01$). Furthermore, we implemented rigorous randomization procedures to balance our experimental groups. This approach to group assignment minimized the potential for confounding variables and strengthened our ability to attribute observed effects to our manipulated conditions.

\subsection{Ecological validity} 

Ecological validity measures how well an experimental setup reflects real-world conditions. Our study environment diverged from typical programming scenarios in some respects, potentially impairing ecological validity, as participants lacked access to an Integrated Development Environment (IDE) or a debugger--common tools used in programming. This design choice, while potentially reducing ecological validity, served to increase internal validity by standardizing the environment across all participants. Moreover, it allowed us to isolate the \change{accuracy} and \change{response time} improvement effects to the program representation and/or the provided hints. Additionally, we selected tasks and code snippets particularly relevant to novice programmers and educational settings. This approach, though not perfectly replicating professional development scenarios, aligned with the experiences of our target population in learning contexts. Thus, we strived to maintain ecological validity while adhering to rigorous experimental controls.

\subsection{External validity}

\looseness-1External validity concerns the ability of our results to generalize to other settings, participants, and measures. We posit that our findings are likely generalizable to similar contexts, i.e., small to medium-sized code snippets and algorithmic tasks of comparable complexity. However, we acknowledge that the effectiveness of these representations and hints may not necessarily extend to large-scale software projects. Additionally, our participant pool, primarily recruited from crowdsourcing platforms, consisted largely of individuals with limited programming experience. This may limit the generalizability of our results to other populations such as computer science students or highly experienced professional programmers. 
\change{Nevertheless, it is important to consider the evolving landscape of software engineering, where diverse user groups -- often untrained in programming -- are increasingly playing central roles in application development~\cite{nocode_quixy}. Our insights align well with this emerging trend.}
\change{Finally,} it is important to note that our study represents one of the first comprehensive investigations into the utility of program representations, hints, and their interactions in the context of bug understanding. As such, \change{our findings offer valuable guidance for developing tools that adapt to programming representations and user expertise.} They also underscore the need for future research to explore the generalizability of these findings across a broader spectrum.

\section{Discussion}
\label{sec:discussion}


In our study, we found that text representations are more beneficial in terms of accuracy than Python representations for individuals with a clear understanding of the algorithmic task. One hypothesis for this possibly counter-intuitive finding is that
Python code is structured and contains implicit beacons (e.g., variable and function names) that aid program comprehension without the need of going through every statement -- a process called \emph{top-down comprehension}~\cite{Bro83,SE84}. 
The text representation forces even seasoned programmers to go through all details from beginning to end, called \emph{bottom-up comprehension}~\cite{SM79,Pen87}, which makes it more likely to spot otherwise hidden bugs. 

\begin{itemize}[leftmargin=0.7cm,topsep=1ex]
    \item[\textbf{H1}:] Python representations trigger top-down comprehension, whereas text representations trigger bottom-up comprehension. The difference between the two will be more pronounced the more experienced programmers are.
\end{itemize}

An alternate hypothesis is that, even though our sample size was large (over 700 participants), the majority of the participants had little self-reported experience with coding or Python and this may also have contributed to their improved \change{accuracy} with text over Python representations. 

\begin{itemize}[leftmargin=0.7cm,topsep=1ex]
    \item[\textbf{H2}:] Text representations aid inexperienced programmers since they strip the algorithmic description from syntactical and technical information arising from the use of a formal programming language.
\end{itemize}

Future work that repeats our experiment in a population of experienced participants with Python can disambiguate between these two hypotheses, as they make very different predictions for experienced and inexperienced participants: the first hypothesis predicts that the more experience with code the participant has, the stronger their prior will be, and the easier it would be to overlook the bug and therefore the gap between text and Python representations will widen. The second hypothesis predicts that the more experience with code the participant has, the more they will benefit from the Python representation in terms of accuracy of finding bugs. 
We conducted a preliminary investigation by analyzing participants divided into three groups based on self-reported programming proficiency: low, medium, and high experience. Our findings revealed that participants across all experience levels performed better with text-based representations than with Python. This effect was significant for those with low \change{(p-value: 0.021, effect-size: $0.452$, difference of means: 0.08)} programming experience. We share the detailed results in our \href{https://github.com/bridge-ai-neuro/HintsCodeText}{GitHub repository} due to space constraints. These initial results suggest \change{evidence towards H2}, i.e., text representations aid inexpierenced programmers.
However, to draw more definitive conclusions, further experiments under controlled conditions using objective and verifiable measures of programming expertise beyond self-reporting are necessary.

We further found that the addition of hints was particularly useful for the Python representation. In these cases, hints improved user \change{accuracy} regardless of whether the participant had a clear or confused understanding. Furthermore, participants with a confused understanding benefited so much from the hint that their \change{accuracy} was statistically indistinguishable from that of the participants with originally clear understanding. This suggests that the lack of algorithmic understanding can be made up using hints. 
Additionally, hints also bridge the gap between the Python and text representations for participants with clear understanding, such that participants who see the program in Python and receive a hint perform at the level of those who see the program in text. This result suggests that hints can be a powerful tool in aiding program debugging in code, when faithful text descriptions are difficult to generate. 

Interestingly, we find that hints have contrasting effects on accuracy and completion time for Python vs.\ natural text representations: hints improve accuracy but not speed for Python representations, and improve speed but not accuracy for text representations. Again, this may be because the Python representation already provides a sufficiently high-level description of the program that can be accessed quickly so the speed is difficult to improve upon. As said in H1, this representation can also make it harder to detect the bugs, which is where hints can help by drawing attention to specific issues in the program. In contrast, text representations require going through the program word-by-word to extract the overall structure and may thus require longer time to integrate the higher-level understanding of the program. While this low-level presentation can be helpful in detecting bugs, manipulating it can be slow, which can be improved by the addition of higher-level hints.

\begin{itemize}[leftmargin=0.7cm,topsep=1ex]
    \item[\textbf{H3}:] 
    Hints alter strategies for program comprehension and bug finding.
\end{itemize}

\looseness-1Lastly, we found that among the $3$ different types of hints--test cases, conceptual, and detailed fix--detailed fixes are generally the most helpful in improving \change{accuracy} across all representation modalities and user types. Additionally, conceptual hints are particularly beneficial for users with a confused understanding when working with Python representations. While we don't find a significant improvement in response accuracy due to test cases, we observe that they lead to the biggest improvement in efficiency (i.e.\ reduced time per accurate response), specifically for text-based representations. Similarly to the benefit of the natural language conceptual hint in the Python representation, here we also observe a benefit when mixing the presentation modalities of the program and the hint. We hypothesize that the mixing of presentation modalities may contribute to a more holistic understanding of the program.

\begin{itemize}[leftmargin=0.7cm,topsep=1ex]
    \item[\textbf{H4}:] Mixing natural text and code representations improves holistic understanding of the program. 
\end{itemize}

Overall, these results suggest a debugging workflow that is personalized to the level of algorithmic understanding of the user: if the user understood the posed algorithmic task, then they benefit most from seeing the program described in natural language. If the user did not understand the algorithmic task from the start, then they most benefit from seeing the program in Python and receiving a conceptual hint. While we focus here on short-term improvements in program understanding, we hope that our results can serve as a starting point for future work that investigates the utility of hints in long-term program understanding over multiple educational sessions. 

\section{Conclusion and Perspectives}

Recent advancements in generative AI have revolutionized programming accessibility, allowing users to solve tasks through various representations, including natural language and pseudo-code. However, these new tools still require users to possess algorithmic thinking and debugging skills. It is currently unclear which type of hint is most helpful and how this depends on the program representation and the user's understanding level. Our study aimed to address this gap by investigating the effectiveness of different hint types across various program representations and user understanding levels.

We conducted a large-scale, crowd-sourced study involving 753 participants to examine the utility of different program representations and hint types in finding and fixing bugs for participants with varying levels of understanding. Specifically, our research focused on three key questions: how program representation affects a user's ability to find and fix bugs, the utility of hints across different program representations, and which hint types are most suitable for various program representations, all in relation to the user's level of understanding of the algorithmic task.

Our findings revealed several important insights:

\begin{enumerate}
    \item Text-based program representations improve \change{accuracy} for users with clear understanding of the algorithmic task.
    \item Hints significantly improved \change{accuracy} for Python representations across both clear and confused understanding groups. For text-based representations, hints  
    \change{reduce response time} for those with clear understanding. Additionally, hints bridged \change{accuracy} gaps between different representations and understanding levels.
    \item Different hint types vary in utility depending on program representation and participant understanding. Detailed fixes generally improved \change{accuracy} most, while \change{test cases} reduce \change{response time} for participants with clear understanding viewing text-based representations.
\end{enumerate}

These results have significant implications for the design of next-generation programming tools. They suggest the potential for personalized support systems that adapt to users' experiences and understanding levels. By tailoring program representations and hint types to individual users, we can enhance their ability to find and fix bugs, ultimately improving their programming skills and \change{reducing response time}. \change{Additionally, for software engineering researchers, our work represents the first study to systematically investigate the intersection of different programming representations and types of hints. Our methodology and study setup provide a blueprint for exploring similar and follow-up research questions, some of which we articulate in Section~\ref{sec:discussion} in the form of explicit hypotheses.}

\section{Data Availability}
\label{sec:data_avail}

The raw data from the human study and scripts to generate all the plots present in the paper are available at \href{https://github.com/bridge-ai-neuro/HintsCodeText}{https://github.com/bridge-ai-neuro/HintsCodeText}.

\section*{Acknowledgements}
The authors would like to thank Norman Peitek, Anna-Maria Maurer, Tung Phung, Ahana Ghosh, Gabriele Merlin, Emin Çelik, and Aritra Mitra for their feedback on our Cloud Research survey workflow. Financial support was provided by the German Research Foundation through Collaborative Research Center TRR 248 (389792660), by the European Union as part of ERC Advanced Grant ``Brains On Code'' (101052182), and by the European Union as part of ERC Starting Grant ``TOPS'' (101039090).


\bibliographystyle{IEEEtran}
\bibliography{IEEEabrv,main}

\end{document}